\documentclass[12pt]{iopart}

\usepackage{iopams}
\usepackage[english]{babel}
\usepackage{graphicx}
\usepackage{cite}
\usepackage{color}
\usepackage{bbm}
\newcommand{\ket}[1]{\vert #1 \rangle}
\newcommand{\bra}[1]{\langle #1 \vert}

\begin{document}

\title[]{Generation of higher dimensional entangled states in quantum Rabi systems}

\author{F. Albarr\'an-Arriagada}
\address{Departamento de F\'isica, Universidad de Santiago de Chile (USACH), 
Avenida Ecuador 3493, 9170124, Santiago, Chile}

\author{G. Alvarado Barrios}
\address{Departamento de F\'isica, Universidad de Santiago de Chile (USACH), 
Avenida Ecuador 3493, 9170124, Santiago, Chile}

\author{F. A. Cardenas-L\'opez}
\address{Departamento de F\'isica, Universidad de Santiago de Chile (USACH), 
Avenida Ecuador 3493, 9170124, Santiago, Chile}
\address{Center for the Development of Nanoscience and Nanotechnology 9170124, Estaci\'on Central, Santiago, Chile}

\author{G. Romero}
\address{Departamento de F\'isica, Universidad de Santiago de Chile (USACH), 
Avenida Ecuador 3493, 9170124, Santiago, Chile}

\author{J. C. Retamal}
\address{Departamento de F\'isica, Universidad de Santiago de Chile (USACH), 
Avenida Ecuador 3493, 9170124, Santiago, Chile}
\address{Center for the Development of Nanoscience and Nanotechnology 9170124, Estaci\'on Central, Santiago, Chile}
\ead{juan.retamal@usach.cl}

\begin{abstract}
We present protocols for the generation of high-dimensional entangled states of anharmonic oscillators by means of coherent manipulation of light-matter systems in the ultrastrong coupling regime. Our protocols consider a pair of ultrastrong coupled qubit-cavity systems, each coupled to an ancilla qubit, and combine classical pulses plus the selection rules imposed by the parity symmetry. We study the robustness of the entangling protocols under dissipative effects. This proposal may have applications within state-of-art circuit quantum electrodynamics.   
\end{abstract}


\maketitle

\section{Introduction}

One of the striking properties of quantum mechanics is quantum entanglement \cite{Horodecki2009}. This is a holistic feature of multipartite quantum systems \cite{Wei2003,Bourennane2004,Carvalho2004,Lopez2008,Li2009} that does not have a classical counterpart. The preparation of highly entangled states is an important research line, since quantum entanglement has proven to be a key resource for different tasks in quantum information \cite{Bennett1993,Bennett2000,Nielsen2010,Li2004}, and quantum metrology \cite{Giovannetti2006,Joo2011,Riedel2010}. For instance, the Dicke and $N00N$ states are a class of multipartite entangled states \cite{Amico2008,Lucke2014} that can be experimentally implemented with photons \cite{Prevedel2009,Wieczorek2009}. Also, several protocols to generate entangled states have been proposed in different physical platforms including optomechanical systems \cite{Vitali2007,Ren2013,Wang2013}, atomic systems \cite{Cirac1994,Ficek2002,Islam2008,Chen2010}, photonics systems~\cite{Mitchell2004,Resch2007,Afek2010,DAmbrosio2013,Barreiro2005}, and superconducting circuits \cite{Merkel2010,Strauch2010,Neely2010,Wang2011,Strauch2012,Su2014}. This quantum technology has experienced a noticeable development in last years with important applications to quantum computing \cite{Devoret2013,Barends2016} and quantum simulations~\cite{Houck2012,Romero2016}. 

In the same way, superconducting circuits and circuit quantum electrodynamics (QED) allow the implementation of light-matter interaction in the ultrastrong~\cite{FornDiaz2010,Niemczyk2010} and deep strong coupling~\cite{Casanova2010,Yoshihara2016} regimes. Here, the light-matter coupling strength is comparable to or larger than the qubit and cavity frequencies, where the equilibrium and nonequilibrium properties of such a system, called quantum Rabi system (QRS), are described by the quantum Rabi model (QRM) \cite{Rabi1936,Braak2011}. The latter has also been implemented in trapped ions~\cite{Pedernales:2015aa,Puebla2016}. The QRM exhibits a discrete parity symmetry ($\mathbb{Z}_2$), which establishes selection rules for state transitions \cite{Forn-Diaz:2016aa,Felicetti2015,Wang2016}. In particular, the QRM has also proven useful for various quantum information tasks~\cite{Nataf2011,Romero2012,Kyaw2015,Felicetti2015,Wang2016,Kyaw2016parity}.       

In this work, we study the conditions under which a coherent manipulation of the Hilbert space of a bipartite QRS can be realized. In particular, we propose schemes to prepare high-dimensional entangled states between two QRSs or polaritons \cite{Rossatto2016,PhysRevLett.112.016401}, each composed by an ultrastrong coupled qubit-cavity system; by means of their resonant interactions with an ancilla two-level system (TLS), and classical pulses. These entangled states are invariant under the exchange of quantum Rabi systems. Specifically, we generate states of the form $|S_{N,M}\rangle=(|NM\rangle+|MN\rangle)/2$, $|D_{N,M}\rangle=(|NN\rangle+|MM\rangle)/2$, and $|T_N\rangle=(|S_{N,0}\rangle+|S_{N-1,1}\rangle+...+|S_{N-N/2,N/2}\rangle)/\mathcal{W}$, with $\mathcal{W}$ a normalization factor. Here, $|N\rangle(|M\rangle)$ is the $N$th ($M$th) excited state for a single QRS. Notice that our system is different from the case of two independent Jaynes-Cummings atoms beyond the rotating-wave approximation \cite{PhysRevA.82.052306}. We study the robustness of the entangling protocols under various loss mechanisms. Our proposal may have applications within circuit QED in the ultrastrong coupling (USC) regime of light-matter interaction. In particular, the above mentioned states may have applications in parameter estimation for sensing magnetic fluxes in a quantum metrology approach. In addition, we could use the individual addressing of different energy transitions of each quantum Rabi system for the quantum simulation of complex spin systems such as Heisenberg interactions between high dimensional spins. We stress that the protocols performance will be much affected in the deep strong coupling regime ~\cite{Casanova2010,Yoshihara2016}, since the QRS spectrum becomes quasiharmonic which is detrimental in the selective excitation exchange between the ancilla TLS and each QRS.

\section{The model and $\mathbb{Z}_2$ symmetry}
\begin{figure}[b]
\centering
\includegraphics[scale=0.2]{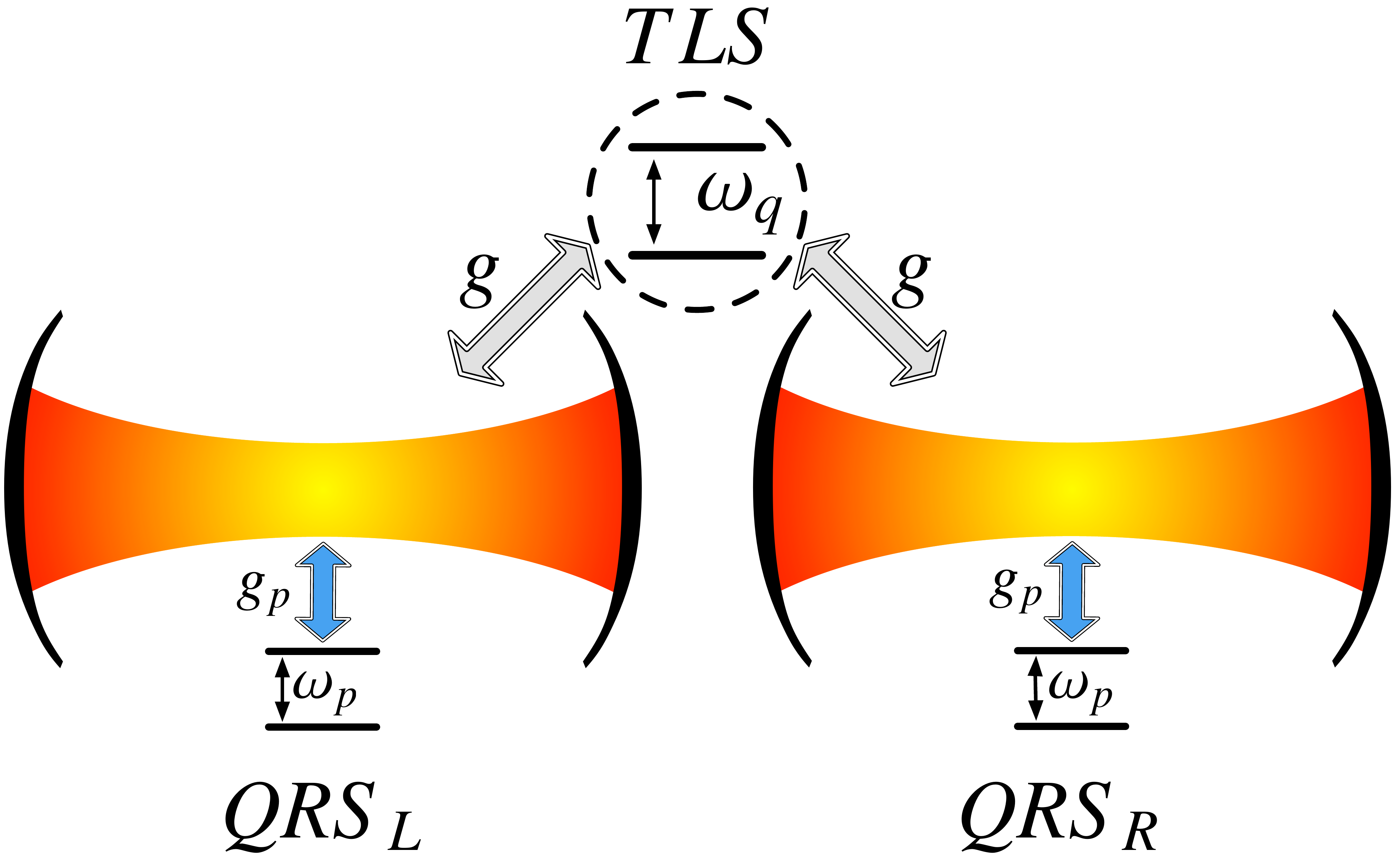}
\caption{Schematics of our model. Two identical quantum Rabi systems interact with a common two-level system ancilla through TLS-cavity coupling.}
\label{figure1}
\end{figure}
Let us consider two identical quantum Rabi systems, each composed by an ultrastrong coupled qubit-cavity system and described by the generalized quantum Rabi model~\cite{Niemczyk2010,Garziano2014,Garziano2016}
\begin{equation}
H_{n,GQRM}=\frac{\hbar}{2}\omega_p\sigma^z_n+\hbar\omega_c a_n^{\dagger}a_n + \hbar g_p[\cos(\theta)\sigma^x_n+\sin(\theta)\sigma^z_n] \left( a_n+a_n^{\dagger} \right),
\label{genpolariton}
\end{equation}
where the index $n=\{L,R\}$ labels the operators associated with the left or right QRS, $a_n$($a_n^{\dagger}$) is the annihilation(creation) operator of the lowest mode of the cavity, and $\sigma_n^x$ and $\sigma_n^z$ are Pauli matrices associated with the qubit. In addition, $\omega_p$, $\omega_c$, $g_p$, and $\theta$ are the qubit frequency gap, cavity frequency, qubit-cavity coupling strength, and the mixing angle, respectively. Selecting $\theta=0$, we obtain the quantum Rabi model \cite{Rabi1936,Braak2011}
\begin{equation}
H_{n,QRM}=\frac{\hbar}{2}\omega_p\sigma^z_n+\hbar\omega_c a_n^{\dagger}a_n + \hbar g_p\sigma^x_n \left( a_n+a_n^{\dagger} \right).
\label{polariton}
\end{equation}

Consider now an ancilla TLS weakly coupled to both quantum Rabi systems through each cavity mode, as depicted in Fig.~\ref{figure1}. The ancilla TLS is the mechanism to introduce excitations into the two QRSs or polaritons, so that it is necessary that it remains disentangled from the states of the quantum Rabi systems at the initial stage of the entangling protocol. Within the strong or ultrastrong coupling regime between the ancilla qubit and both polaritons, they will become a whole system whose ground state is entangled. In this case the separability condition of the initial state for the whole system will not be longer valid. This situation is modeled by the Hamiltonian
\begin{equation}
H=H_o + \hbar g\sigma_q^x\left[\left( a_L+a_L^{\dagger}\right) + \left( a_R+a_R^{\dagger}\right) \right],
\label{system}
\end{equation}
 where $H_o$ is the free Hamiltonian that reads
\begin{equation}
H_o=\sum_{n=L,R}H_{n,QRM}+H_{\rm TLS},
\label{Ho}
\end{equation}
and $H_{\rm TLS}=(\hbar\omega_q/2)\sigma_q^z$. Here, $\sigma_q^x$, $\sigma_q^z$, and $\omega_q$, are Pauli matrices and the frequency gap associated with the ancilla TLS, respectively. Also, $g$ stands for the coupling strength between the TLS and each QRS. 

\begin{figure}[t]
	\centering
	\includegraphics[scale=0.2]{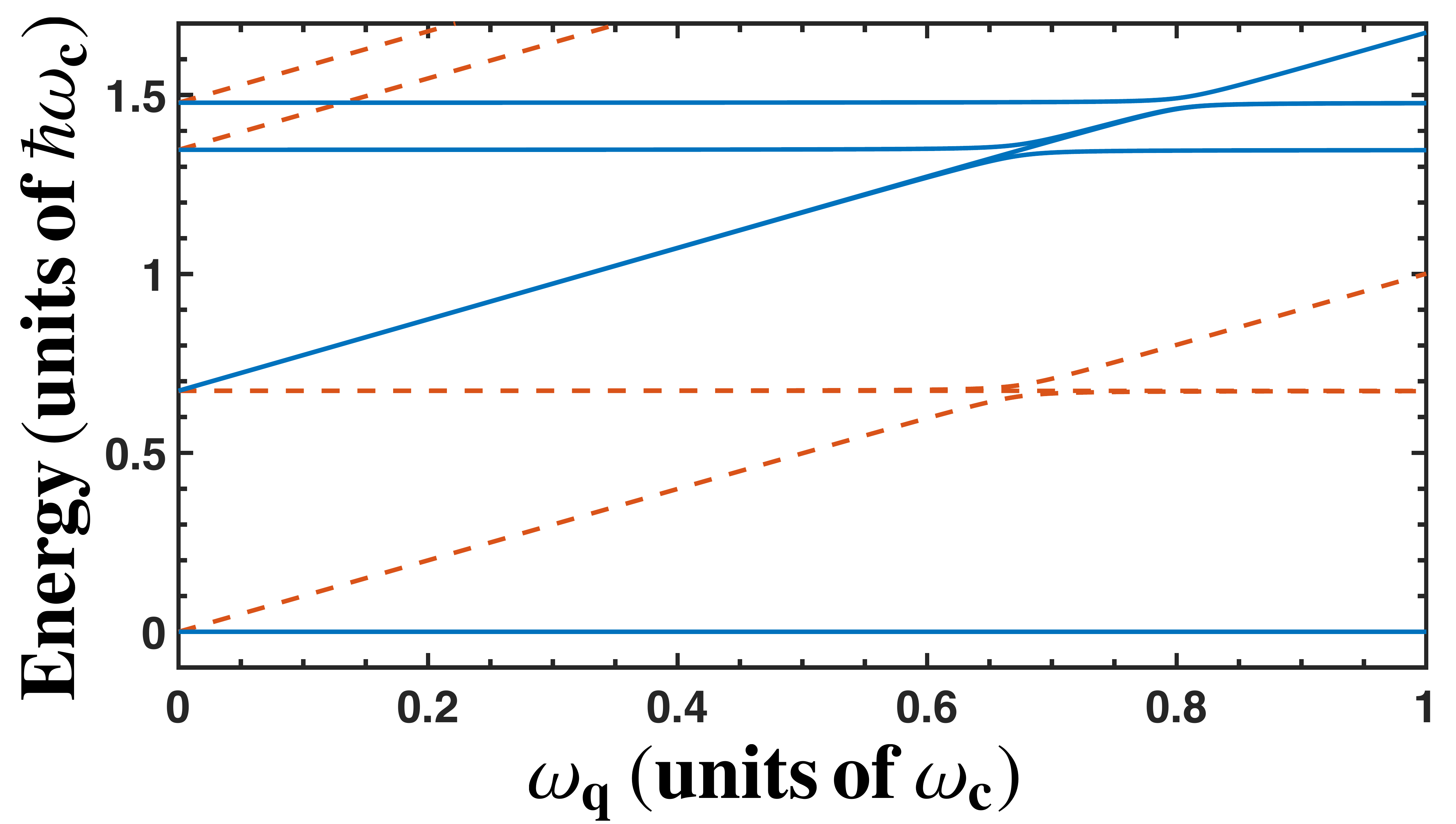}
	\caption{Energy differences with respect to the ground state of the total Hamiltonian (\ref{system}) as function of the TLS frequency $\omega_q$. In this simulation we considered QRS parameters $\omega_p=0.8\omega_c$ $g_p=0.5\omega_c$ and $g=10^{-2}\omega_c$. The blue (continuous) lines stand for states with parity $p=+1$ and red (dashed) lines for states with parity $p=-1$.}
	\label{fig:figure2}
\end{figure}

It is instructive to show that the global system exhibits a discrete parity symmetry, that is, the Hamiltonian (\ref{system}) is invariant under the change $\sigma^x_n\to-\sigma^x_n$, $\sigma_q^x\to-\sigma_q^x$, and $a_n+a^{\dag}_n\to-(a_n+a^{\dag}_n)$. Hence, there exists an operator $\mathcal{P}=-e^{i\pi (a^{\dagger}_La_L+a^{\dagger}_Ra_R)}\sigma_L^z\sigma_R^z\sigma_q^z$ that commutes with the Hamiltonian (\ref{system}). Since $[H,\mathcal{P}]=0$, there must be a common basis $\{|\varphi_j\rangle\}$ that simultaneously diagonalizes $H$ and $\mathcal{P}$ such that, $\mathcal{P}|\varphi_j\rangle=p|\varphi_j\rangle$, and $H|\varphi_j\rangle=\varepsilon_j|\varphi_j\rangle$. The parity symmetry separates the Hilbert space into two sub-spaces, one of them with parity $p=+1$ and the other with parity $p=-1$. Figure \ref{fig:figure2} shows the spectrum of the total system given by Eq.~(\ref{system}) as a function of the TLS frequency $\omega_q$. Blue (continuous) lines stand for states with parity $p=+1$, and red (dashed) lines for states with parity $p=-1$. Notice that the shift of the TLS frequency does not introduce any parity breaking mechanism. At the same time, the spectrum shows avoided level crossings between states that belong to the same parity subspace \cite{Kyaw2016parity}. 

\begin{figure}[t]
	\centering
	\includegraphics[scale=0.2]{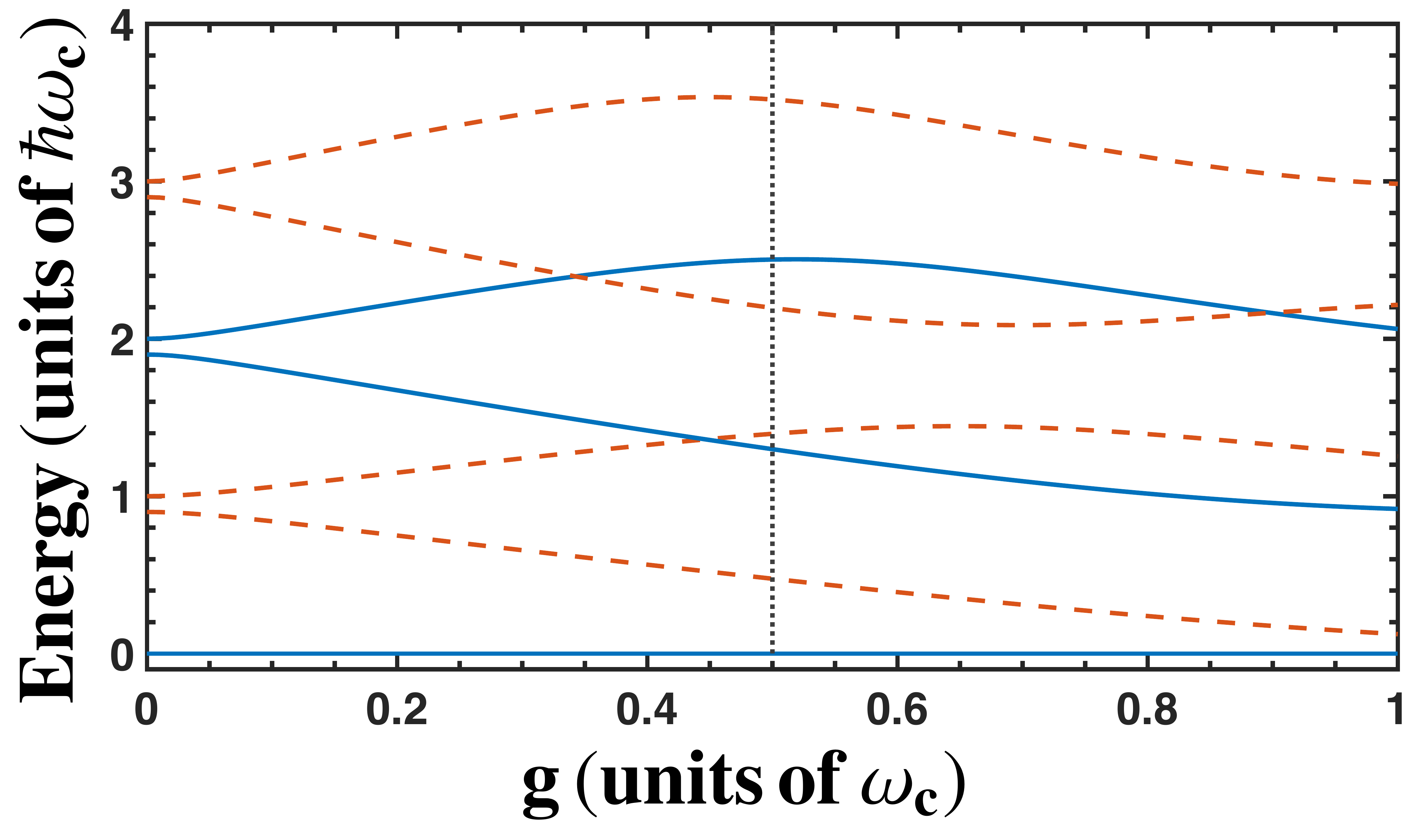}
	\caption{Energy differences with respect to the ground state of the Hamiltonian (\ref{polariton}) as a function of the coupling strength $g_p$ and $\omega_p=0.8\omega_c$. Blue (continuous) lines represent states with parity $\pi_{n}=+1$ and red (dashed) lines for states with parity $\pi_{n}=-1$ associated with the parity operator $\Pi_n= -e^{i\pi a^{\dagger}_{n}a_{n}}\sigma_{n}^z$. The dashed vertical line corresponds to the coupling strength $g_p=0.5\omega_{\rm cav}$ that we use in our numerical calculations.}
	\label{fig:figure3}
\end{figure}

The above symmetry has its origin in the intrinsic $\mathbb{Z}_2$ symmetry for a single QRS, described by $H_{n,QRM}$ in Eq.~(\ref{polariton}), and the parity operator $\Pi_{n}=-e^{i\pi a^{\dagger}_{n}a_{n}}\sigma_{n}^z$. In fact, $\mathcal{P}=-\Pi_L\otimes\Pi_R\otimes\sigma_q^z$. Fig.~\ref{fig:figure3} shows the spectrum of $H_{n,QRM}$, where blue (solid) lines stand for states with eigenvalue $\pi_{n}=+1$ and red (dashed) lines for $\pi_{n}=-1$ associated with the operator $\Pi_{n}$. The vertical (dotted) line is placed at $g=0.5\omega_c$ which is used in Fig.\ref{fig:figure2}.

As a result of the parity symmetry, the system described by the Hamiltonian (\ref{system}) features selection rules that can be explained as follows. States with same parity can be connected only by operators that preserve the parity symmetry such as $\sigma^z_n$ or $\sigma_q^z$ ($[\mathcal{P},\sigma^z_n(\sigma_q^z)]=0$). The latter corresponds exactly to the case of shifting the TLS frequency $\omega_q$, and gives rise to the avoided level crossings that appear in Fig.~\ref{fig:figure2}. States with different parity can only be connected by interactions that break the symmetry, for instance, by means of driving the cavities through their field quadratures $X_n=a^{\dagger}_n+a_n$, or a qubit driving proportional to $\sigma^{x}_n$ or $\sigma_q^{x}$.  These rules will be essential for developing our entangling protocols as we will describe in section \ref{sec:IV}. 

In order to further explore the parity symmetry, it is convenient to write the Hamiltonian (\ref{system}) in the basis of product states among the two quantum Rabi systems $\{\ket{j_L}\otimes\ket{j_R}\}$, and the pseudo-spin $\{\ket{\uparrow},\ket{\downarrow}\}$, where the free Hamiltonian $H_o$ is diagonal (dressed basis)
\begin{eqnarray}
H_{n,QRM}|j_n\rangle &&=\hbar\omega_j|j_n\rangle,\nonumber\\
H_{TLS}\ket{\uparrow}(\ket{\downarrow})&&=+\frac{\hbar\omega_q}{2} \ket{\uparrow}(-\ket{\downarrow}).
\end{eqnarray}
Here, the index $j=0,1,2,...$, labels the energy states for a single QRS, with eigenenergies $\hbar\omega_j$. Notice that the eigenbasis $\{|j_n\rangle\}$ represent polaritonic states, this means a hybrid light-matter eigenstate of the quantum Rabi system \cite{Rossatto2016,PhysRevLett.112.016401}. Now, the field quadrature $X_n=a_n^{\dagger}+a_n$ reads
\begin{equation}
X_n=\sum\limits_{k,j>k}\chi_{jk}\left( |j_n\rangle \langle k_n| + |k_n\rangle \langle j_n| \right), 
\label{Xp}
\end{equation}
where $\chi_{jk}=\langle j_n|X_n|k_n\rangle $, remember that $n=\{L,R\}$. For all states $|j_n\rangle$ and $|k_n\rangle$ belonging to the same parity subspace $\chi_{jk}=0$. Therefore, the total Hamiltonian $(\ref{system})$ reads
\begin{eqnarray}
H=&&\hbar\sum\limits_{j=0}^{\infty}\sum\limits_{n={L,R}}\omega_j|j_n\rangle\langle j_n|+\frac{\hbar\omega_q}{2}\sigma_q^z\nonumber\\
&&+\hbar g\sigma^x \sum\limits_{k,j>k=0}^{\infty}\sum\limits_{n=L,R}\chi_{jk}\left( |j_n\rangle \langle k_n| + |k_n\rangle \langle j_n| \right).
\label{Hpolbase}
\end{eqnarray}
The above Hamiltonian in the interaction picture with respect to $H_o$, is given by
\begin{eqnarray}
H_I(t)=&&\hbar\sigma_q^+\sum\limits_{k,j>k=0}^{\infty}\Omega_{jk}\left(  e^{i\delta_{jk}t}B_{jk}^{\dagger} + e^{i\Delta_{jk}t}B_{jk}\right)\nonumber\\
&&+ \hbar\sigma_q^-\sum\limits_{k,j>k=0}^{\infty}\Omega_{jk}\left(  e^{-i\Delta_{jk}t}B_{jk}^{\dagger} + e^{-i\delta_{jk}t}B_{jk}\right),
\label{Hinteraction}
\end{eqnarray}
where $\Delta_{jk}=\omega_q-(\omega_j-\omega_k)$, $\delta_{jk}=\omega_q+(\omega_j-\omega_k)$ are the frequency difference and sum of the TLS frequency gap and the $(j,k)$ QRS transition respectively, $\Omega_{jk}=\sqrt{2}g\,\chi_{jk}$ is the effective coupling strength between the TLS and the $jk$ QRS transition, and we define $B^{\dagger}_{jk}=(\ket{j_L}\bra{k_L}+\ket{j_R}\bra{k_R})/\sqrt{2}$ and $B_{jk}=(\ket{k_L}\bra{j_L}+\ket{k_R}\bra{j_R})/\sqrt{2}$, as the collective raising and lowering operators for identical quantum Rabi systems.

In the next section we introduce the effective Hamiltonians for three different regimes which can be derived from Hamiltonian (\ref{Hinteraction}), that we shall use in the entangling protocols. Hereafter, we refer to these interactions as SWAP, TLS rotation, and QRS rotation. We stress that all numerical calculations have been performed via the \textit{ab initio} Hamiltonian (\ref{system}).
\section{Resonant Interactions}
\subsection{SWAP Interaction}
\begin{figure}[t]
\centering
\includegraphics[scale=0.2]{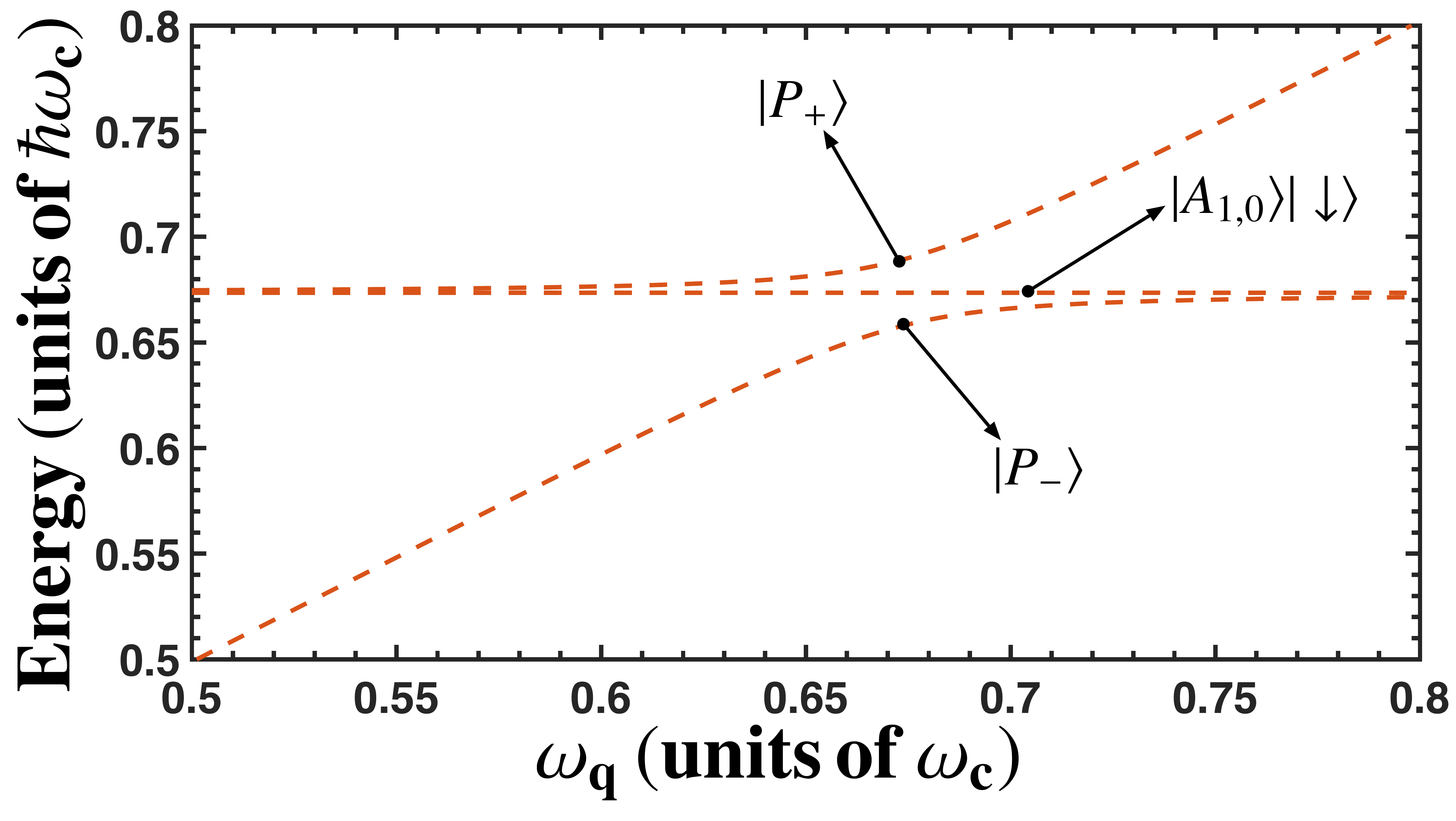}
\caption{Zoom of the first avoided level crossing in Fig.~\ref{fig:figure2}. The maximally entangled states $\ket{P_{\pm}}$ and the antisymmetric state $\ket{A_{1,0}}\ket{\downarrow}$ are shown at the resonance condition $\omega_q=\omega_1-\omega_0$.}
\label{fig:figure4}
\end{figure}

Selecting the gap of the TLS to a specific transition of both quantum Rabi systems, that is, $\Delta_{jk}=\omega_q-\omega_j+\omega_k=0$, and choosing $\Omega_{jk}\ll\delta_{jk}$ allow us to neglect the fast oscillating terms in Eq.~(\ref{Hinteraction}), so that the effective Hamiltonian is given by
\begin{equation}
	H_{\rm SWAP}=\hbar \Omega_{jk}\left( \sigma_q^+B_{jk} + \sigma_q^-B_{jk}^{\dagger}\right).
	\label{SWAP}
\end{equation}
This interaction allows us to swap excitations among the TLS and both QRSs. The quantum dynamics given by $U^{S}_{jk}(t)=\exp(-iH_{\rm SWAP}t/\hbar)$ leads to  
\begin{eqnarray}
|k_Lk_R\rangle|\downarrow\rangle&&\rightarrow |k_Lk_R\rangle|\downarrow\rangle \nonumber\\
|k_Lk_R\rangle|\uparrow\rangle&&\rightarrow \cos(\Omega_{jk}t)|k_Lk_R\rangle|\uparrow\rangle-i\sin(\Omega_{jk}t)|S_{j,k}\rangle|\downarrow\rangle\nonumber\\
|S_{j,k}\rangle|\uparrow\rangle&&\rightarrow \cos(\Omega_{jk}t)|S_{j,k}\rangle|\uparrow\rangle-i\sin(\Omega_{jk}t)|j_Lj_R\rangle|\downarrow\rangle\nonumber\\
|S_{l,k}\rangle|\uparrow\rangle&&\rightarrow \cos\left( \frac{\Omega_{jk}}{\sqrt{2}}t\right) |S_{l,k}\rangle|\uparrow\rangle- i\sin\left( \frac{\Omega_{jk}}{\sqrt{2}}t\right)|S_{l,j}\rangle|\downarrow\rangle\nonumber\\
|j_Lj_R\rangle|\uparrow\rangle&&\rightarrow |j_Lj_R\rangle|\uparrow\rangle, \nonumber\\
\label{resonance}
\end{eqnarray}
where $|S_{j,k}\rangle=(|k_L\rangle|j_R\rangle+|j_L\rangle|k_R\rangle)/\sqrt{2}$ denotes a symmetric state for QRSs.  The antisymmetric states, $|A_{j,k}\rangle=(|k_L\rangle|j_R\rangle-|j_L\rangle|k_R\rangle)/\sqrt{2}$, do not appear in the dynamics because the interaction only couples the TLS states to symmetric states of the QRSs, see the straight line in Fig.~\ref{fig:figure4}, which corresponds to the state $\ket{A_{1,0}}\ket{\downarrow}$. It is also shown that at the avoided level crossing, the TLS and both QRSs hybridize to form maximally entangled states $\ket{P_{\pm}}=(\ket{0_L0_R}\ket{\uparrow}\pm\ket{S_{1,0}}\ket{\downarrow})/\sqrt{2}$.

\subsection{TLS Rotation}
Local operations on the TLS  are necessary to generate entangled states. A rotation can be achieved by switching on a classical driving acting on the TLS, 
\begin{equation}
H^D_{TLS}=\hbar\Omega_{\mu} \cos\left( \mu t + \phi\right) \sigma_q^x.
\label{rotationTLS}
\end{equation}
In this case the total Hamiltonian is $\tilde{H}=H+H^D_{TLS}$, where $H$ is given by Eq.~(\ref{Hpolbase}). In the interaction picture we obtain
\begin{eqnarray}
V^D_{TLS}=&&\sigma_q^+\sum\limits_{k,j>k=0}^{\infty}\hbar\Omega_{jk}\left(  e^{i\delta_{jk}t}B_{jk}^{\dagger} + e^{i\Delta_{jk}t}B_{jk}\right)\nonumber\\
&&+ \frac{\hbar\Omega_\mu}{2}\sigma_q^+\left[e^{i(\omega_q+\mu)t+i\phi} + e^{i(\omega_q-\mu)t-i\phi}\right]  + {\rm H.c.}
\end{eqnarray}
We choose the driving frequency on resonance with the TLS gap and off resonance with any allowed transition on both QRSs, that is $\mu=\omega_q\ll(\omega_j-\omega_k)$. In addition, given that $\Omega_{jk},\,\Omega_\mu\ll\omega_q$, we can perform a rotating-wave approximation obtaining the following Hamiltonian,
\begin{equation}
H^R_{TLS}(\phi)=\frac{\hbar\Omega_\mu}{2}\left( e^{-i\phi}\sigma_q^++e^{i\phi}\sigma_q^-\right).
\end{equation}
The TLS evolves according to the unitary operation $U^{TLS}(t,\phi)=\exp(-iH^R_{TLS}(\phi)t/\hbar)$ leading to
\begin{equation}
|\downarrow\rangle\rightarrow \cos\left( \frac{\Omega_\mu}{2}t\right) |\downarrow\rangle - ie^{-i\phi} \sin\left( \frac{\Omega_\mu}{2}t\right)|\uparrow\rangle
\label{rotationG}
\end{equation}
\begin{equation}
|\uparrow\rangle\rightarrow \cos\left( \frac{\Omega_\mu}{2}t\right) |\uparrow\rangle - ie^{i\phi} \sin\left( \frac{\Omega_\mu}{2}t\right)|\downarrow\rangle,
\label{rotationE}
\end{equation}

\subsection{QRS Rotation}
Local rotations acting on a QRS can be achieved by means of a classical driving on the cavity, that is
\begin{equation}
H^D_{n,QRS}=\hbar\Omega_{\nu} \cos\left( \nu t + \phi\right) (a_n^{\dagger}+a_n).
\label{rotation}
\end{equation}
In this case, the system is governed by $\tilde{H}=H+H^D_{QRS(n)}$, where $H$ is given by Eq.~(\ref{Hpolbase}). In the interaction picture we obtain

\begin{eqnarray}
V^D_{n,QRS}=&&\sigma_q^+\sum\limits_{k,j>k=0}^{\infty}\hbar\Omega_{jk}\left(  e^{i\delta_{jk}t}B_{jk}^{\dagger} + e^{i\Delta_{jk}t}B_{jk}\right)\nonumber\\
&&+\sum\limits_{k,j>k=0}^{\infty} \frac{\hbar\Omega_\nu}{2}\chi_{jk}|j_n\rangle\langle k_n|\left[e^{i(\omega_j-\omega_k+\nu)t+i\phi} + e^{i(\omega_j-\omega_k-\nu)t-i\phi}\right] \nonumber\\
&&+ {\rm H.c.}
\end{eqnarray}
Choosing the driving frequency on resonance with the $(j,k)$ QRS transition, and off resonant with the TLS frequency gap and any other $(l,m)$ QRS transition, that is $\nu=(\omega_j-\omega_k)\gg\omega_q$, and given that $\Omega_{jk},\,\chi_{jk}\Omega_\nu\ll\nu$, we can use a rotating-wave approximation obtaining the following Hamiltonian,
\begin{eqnarray}
&&H^R_{n,QRS}(\phi)=\frac{\hbar\Omega_\nu}{2}\chi_{jk}\left( e^{-i\phi}|j_n\rangle\langle k_n|+e^{i\phi}|k_n\rangle\langle j_n|\right).
\end{eqnarray}
The QRS evolves according to the unitary operation $U^{QRS}_n(t,\phi)=\exp{(-iH^R_{n,QRS}(\phi)t/\hbar)}$ leading to
\begin{eqnarray}
|k\rangle&&\rightarrow \cos\left( \chi_{jk}\frac{\Omega_\nu}{2}t\right) |k\rangle - ie^{-i\phi} \sin\left( \chi_{jk}\frac{\Omega_\nu}{2}t\right)|j\rangle\\
|j\rangle&&\rightarrow \cos\left( \chi_{jk}\frac{\Omega_\nu}{2}t\right) |j\rangle - ie^{i\phi} \sin\left( \chi_{jk}\frac{\Omega_\nu}{2}t\right)|k\rangle,
\label{rotationJ}
\end{eqnarray}

\section{Entangled States Protocol}
\label{sec:IV}
\subsection{$|S_{N,M}\rangle$ and $|D_{N,M}\rangle$ State Protocol}
\begin{figure}[t]
	\centering
	\includegraphics[scale=0.2]{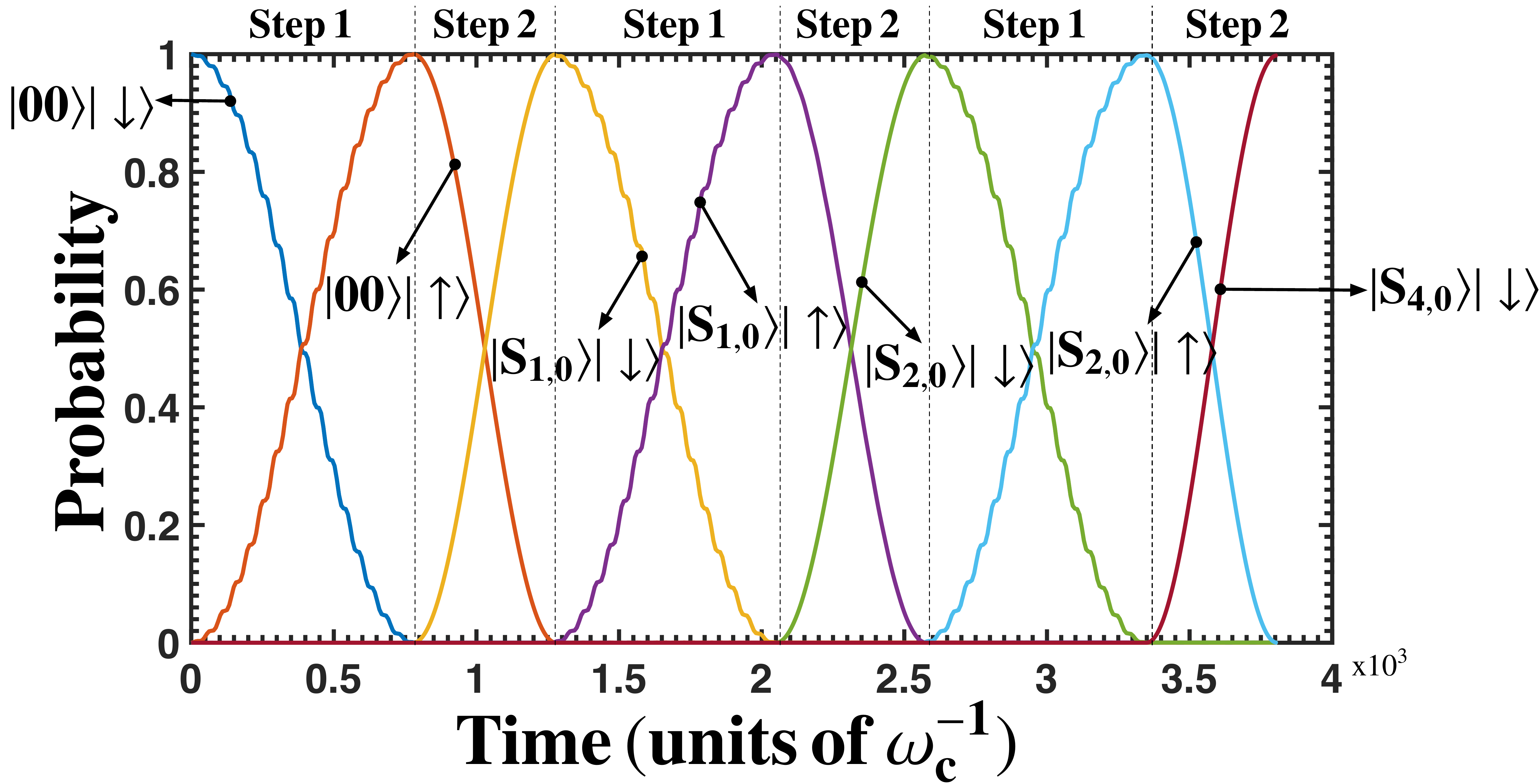}
	\caption{Generation of a $|S_{4,0}\rangle$ state. The QRS parameters are $\omega_p=0.8\omega_c$, $g_p=0.5\omega_c$, and the TLS rotation parameters are $\Omega=0.004\omega_c$, $\mu=0.1(\tilde{\omega}_1-\tilde{\omega}_0)$, and $g=0.002\omega_c$.}
	\label{fig:figure5}
\end{figure}

In what follows we discuss the generation of $|S_{N,M}\rangle$ and $|D_{N,M}\rangle$ states by using the three resonant interactions given in the previous section. These protocols have two stages, the first one generates a $|S_{N,0}\rangle$ state, that corresponds to a $N00N$ state. We repeat this process in order to reach higher $N$.  The second stage uses QRS rotations to obtain the $|S_{N,M}\rangle$ or $|D_{N,M}\rangle$ state. Let us start in the ground state of the Hamiltonian (\ref{system}), which corresponds to both QRSs and the TLS in their ground states
\begin{equation}
|\psi_o\rangle=|00\rangle|\downarrow\rangle.
\label{inistate}
\end{equation}
Hereafter, we disregard the subindexes $L$ and $R$ for the QRS states.\\
\textit{\textbf{Step 1:}} TLS rotation, given by $U_{TLS}(\pi/\Omega_\mu,0)$,
\begin{equation}
U^{TLS}(\pi/\Omega_\mu,0)|00\rangle|\downarrow\rangle=-i|00\rangle|\uparrow\rangle.
\label{step0}
\end{equation}
\textit{\textbf{Step 2:}} Now we perform a $\pi/2$ swapping to the transition $(k,0)$ for both QRSs, that is, $U_{k0}^S(\pi/(2\Omega_{k0}))$, where the QRS state $|k\rangle$ has parity $\pi_n=-1$, opposite to level $|0\rangle$. In this case we obtain  
\begin{equation}
U_{k0}^S(\pi/(2\Omega_{k0})|00\rangle|\uparrow\rangle =-\sqrt{\frac{1}{2}}\left( |k0\rangle+|0k\rangle\right) |\downarrow\rangle,
\end{equation}
which is a $|S_{k,0}\rangle$ state with parity $p=-1$. If we need a $|S_{j,0}\rangle$ state with parity $p=+1$, then we repeat these two steps as follows\\
\textit{\textbf{Step 1 ($2^{nd}$ iteration):}}
\begin{equation}
-U^{TLS}(\pi/\Omega_\mu,0)\sqrt{\frac{1}{2}}\left( |k0\rangle+|0k\rangle\right) |\downarrow\rangle=i\sqrt{\frac{1}{2}}\left( |k0\rangle+|0k\rangle\right) |\uparrow\rangle
\end{equation}
\textit{\textbf{Step 2 ($2^{nd}$ iteration):}} Now, the swapping is performed to the $(j,k)$ transition, that is, $U_{jk}^S(\pi/(\sqrt{2}\Omega_{k0}))$
\begin{equation}
iU_{k0}^S(\pi/2\Omega_{k0})\sqrt{\frac{1}{2}}\left( |k0\rangle+|0k\rangle\right) |\uparrow\rangle=\sqrt{\frac{1}{2}}\left( |j0\rangle+|0j\rangle\right) |\downarrow\rangle,
\end{equation}
thus obtaining a $|S_{j,0}\rangle$ state with parity $p=+1$. In order to prepare a $|S_{N,0}\rangle$ state with high $N$, we need to take into account that physical implementations of our protocol could have limitations when tuning to high frequency values for the TLS gap. For instance, if we consider a circuit QED implementation of our protocols, the ancilla qubit may have a minimum frequency gap of about $1$~[GHz] limited by the thermal noise, and a maximum frequency gap of about $16$~[GHz] limited by the measuring instruments such as microwaves amplifiers \cite{PrivateComm_Pol}. To overcome this problem, we must repeat the two previous steps to obtain an arbitrary $|S_{N,0}\rangle$ state for both QRSs. If we need a $|S_{N,0}\rangle$ state with parity $p=+1(-1)$, then an odd(even) number of repetitions have to be performed. As an example consider the generation of $|S_{4,0}\rangle$ state. Let us start in the state $|00\rangle|\downarrow\rangle$ and apply three repetitions of the two steps for transitions $(1,0)$, $(2,1)$ and $(4,2)$ as follow
\begin{equation}
U^S_{10}(\pi/(2\Omega_{10}))U^{TLS}(\pi/\Omega_\mu,0)|00\rangle|\downarrow\rangle\rightarrow-|S_{10}\rangle|\downarrow\rangle
\end{equation} 
\begin{equation}
-U^S_{21}(\pi/(\sqrt{2}\Omega_{21}))U^{TLS}(\pi/\Omega_\mu,0)|S_{10}\rangle|\downarrow\rangle\rightarrow|S_{20}\rangle|\downarrow\rangle
\end{equation} 
\begin{equation}
U^S_{42}(\pi/(\sqrt{2}\Omega_{42}))U^{TLS}(\pi/\Omega_\mu,0)|S_{20}\rangle|\downarrow\rangle\rightarrow-|S_{40}\rangle|\downarrow\rangle.
\end{equation} 
The numerical calculation of this process is shown in Fig. \ref{fig:figure5}, where we use $\omega_p=0.8\omega_c$, $g_p=0.5\omega_c$, $\Omega_\mu=0.004\omega_c$, $\mu=0.1(\omega_1-\omega_0)$, and $g=0.002\omega_c$.

The state $|S_{N,M}\rangle$ can be obtained by applying a rotation $U^{QRS}$ to the state $|S_{N,0}\rangle$ with driving frequency $\nu=\omega_M-\omega_0$. This interaction applies on both QRSs. The state $|D_{N,M}\rangle$ can be obtained by applying a rotation $U^{QRS}$ to the state $|S_{N,M}\rangle$ with driving frequency $\nu=\omega_N-\omega_M$. This interaction applies on the left or right QRS.

\subsection{$|T_N\rangle$ Protocol}
In order to generate the state $|T_N\rangle=(|S_{N,0}\rangle+|S_{N-1,1}\rangle+...+|S_{N-N/2,N/2}\rangle)/\mathcal{W}$, with $\mathcal{W}$ a normalization factor, we need to superpose all $|S_{J,K}\rangle$ states with $J+K=N$. However, the states that participate in the linear superposition do not have the same parity according to the operator $\mathcal{P}=-e^{i\pi (a^{\dagger}_La_L+a^{\dagger}_Ra_R)}\sigma_L^z\sigma_R^z\sigma_q^z$. For example, if we consider the parameters used in Fig.~\ref{fig:figure2}, the state $|S_{4,0}\rangle\ket{\downarrow}$ has parity $p=-1$, and the state $|S_{3,1}\rangle\ket{\downarrow}$ has parity $p=+1$. One possibility of accessing both types of states is to break the $\mathbb{Z}_2$ symmetry by letting the mixing angle be $\theta\ne0$ in Eq.~(\ref{genpolariton}).  For instance, by choosing $\theta=\pi/4$ we obtain the generalized QRM
\begin{equation}
H_{n,GQRM}=\frac{1}{2}\hbar\omega_p\sigma^z_n+\hbar \omega_ca^{\dagger}_na_n+ \frac{\hbar g_p}{\sqrt{2}}(\sigma_n^x+\sigma_n^z)(a^{\dagger}_n+a_n).
\label{QRSsb}
\end{equation} 
\begin{figure}[t]
	\centering
	\includegraphics[scale=0.2]{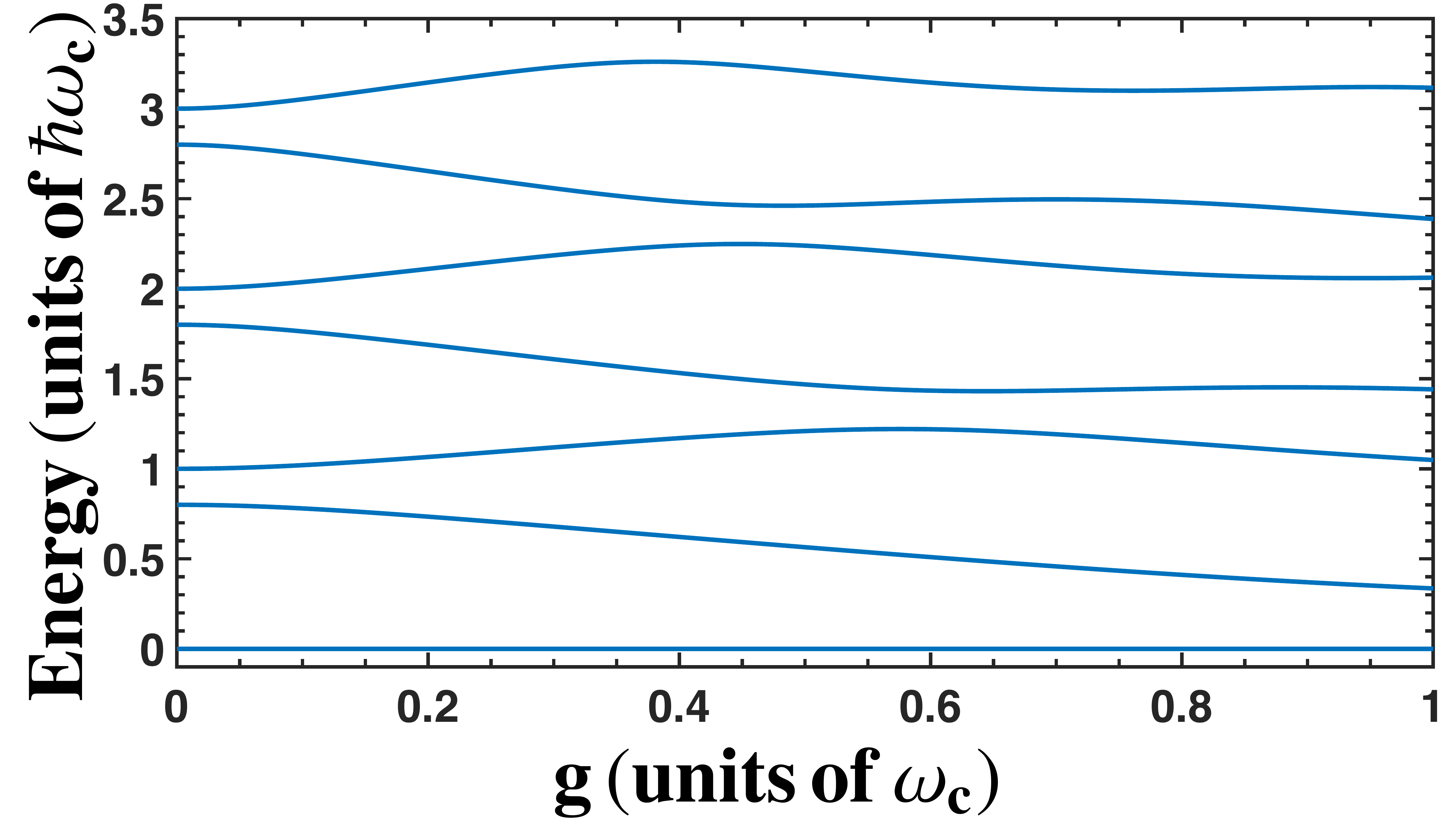}
	\caption{Energy differences with respect to the ground state of the Hamiltonian (\ref{QRSsb}) as a function of the coupling strength $g_p$. In this simulation we consider $\omega_p=0.8\omega_c$.}
	\label{fig:figure6}
\end{figure}

\begin{figure}[b]
	\centering
	\includegraphics[scale=0.2]{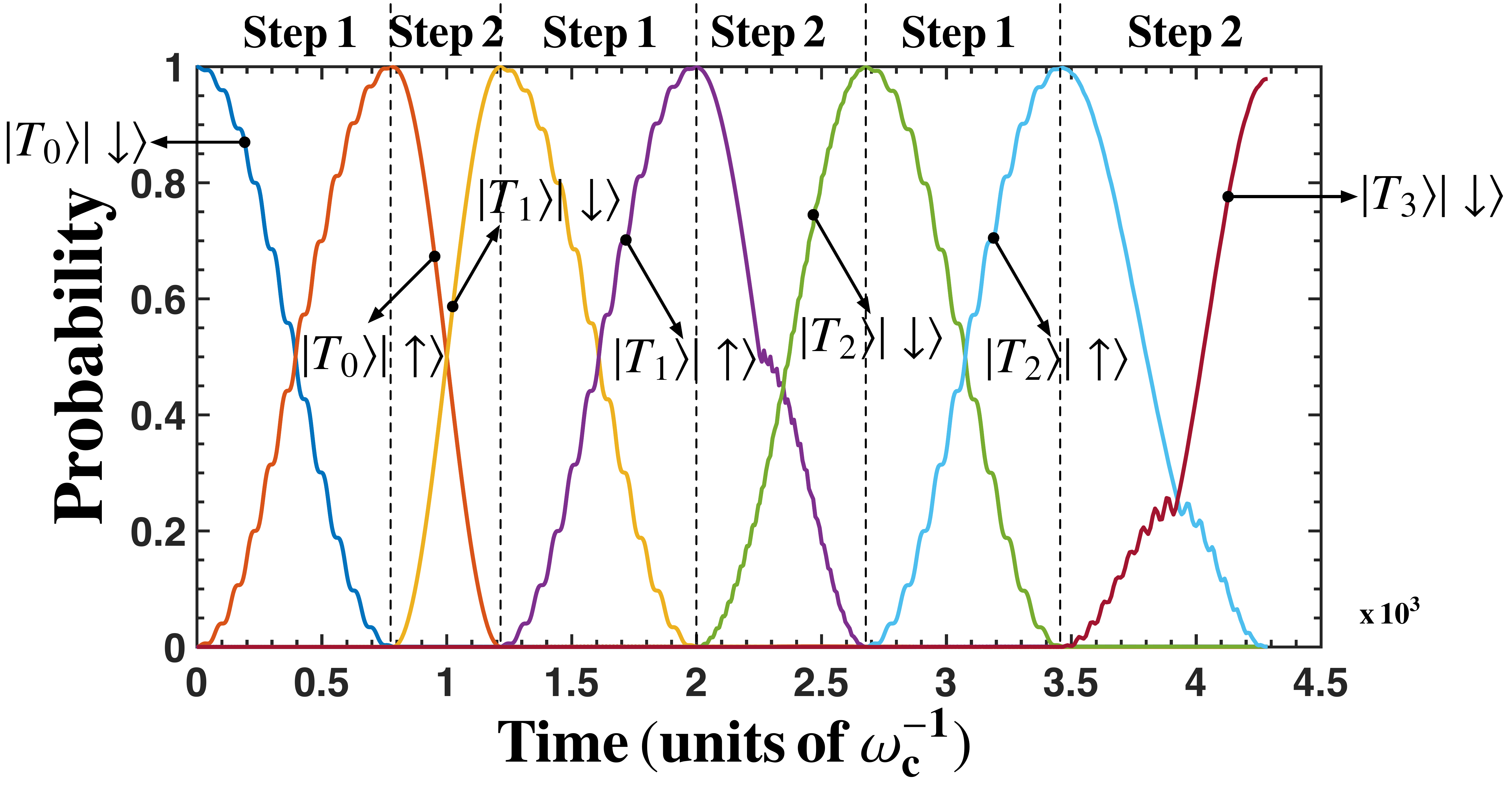}
	\caption{Generation of $|T_N\rangle$ state with $N=3$. The QRS parameters are $\omega_p=0.8\omega_c$, $g_p=0.5\omega_c$, and the TLS rotation parameters are $\Omega=0.004\omega_c$, $\mu=0.1(\tilde{\omega}_1-\tilde{\omega}_0)$, and $g=0.002\omega_c$.}
	\label{fig:figure7}
\end{figure}
In this case, the dynamics given by the SWAP interaction, TLS rotation, and QRS rotation apply now to transitions between states belonging to the spectrum of Hamiltonian (\ref{QRSsb}).

Let us consider the generation of the state $|T_3\rangle$ starting from the ground state of the Hamiltonian $(\ref{QRSsb})$, that is, $|00\rangle|\downarrow\rangle$. In this case, we apply the following steps \\
\textit{\textbf{Step 1:}} TLS rotation, given by $U^{TLS}(\pi/\Omega_\mu,0)$,
\begin{equation}
|00\rangle|\downarrow\rangle\rightarrow -i|00\rangle|\uparrow\rangle.
\end{equation}
\textit{\textbf{Step 2:}} SWAP interaction given by $U^{S}_{10}(\pi/(2\Omega_{10}))$,
\begin{equation}
-i|00\rangle|\uparrow\rangle\rightarrow-|S_{1,0}\rangle|\downarrow\rangle=-|T_1\rangle|\downarrow\rangle.
\end{equation}
\textit{\textbf{Step 1 ($2^{nd}$ iteration):}} TLS rotation $U^{TLS}(\pi/\Omega_\mu,0)$,
\begin{equation}
-|T_1\rangle|\downarrow\rangle\rightarrow i|T_1\rangle|\uparrow\rangle.
\end{equation}
\textit{\textbf{Step 2 ($2^{nd}$ iteration):}} A set of two SWAP interactions given by $\mathcal{U}=U^{S}_{10}(\pi/(2\Omega_{10}))U^{S}_{20}(\pi/(2\sqrt{2}\Omega_{20}))$,
\begin{eqnarray}
&&iU^{S}_{10}(\pi/(2\Omega_{10}))U^{S}_{20}(\pi/(2\sqrt{2}\Omega_{20}))|T_1\rangle|\uparrow\rangle\nonumber\\
&&=iU^{S}_{10}(\pi/(2\Omega_{10}))\frac{1}{\sqrt(2)}(|S_{1,0}\rangle|\uparrow\rangle -i|S_{2,0}\rangle|\downarrow\rangle)\nonumber\\
&&=\frac{1}{\sqrt(2)}(|11\rangle|\downarrow\rangle +|S_{2,0}\rangle|\downarrow\rangle)=|T_2\rangle|\downarrow\rangle
\end{eqnarray}
\textit{\textbf{Step 1 ($3^{rd}$ iteration):}} TLS rotation $U^{TLS}(\pi/\Omega_\mu,0)$,
\begin{equation}
|T_2\rangle|\downarrow\rangle\rightarrow -i|T_2\rangle|\uparrow\rangle.
\end{equation}
\textit{\textbf{Step 2 ($3^{rd}$ iteration):}} A set of two SWAP interactions given by $\mathcal{U}=U^{S}_{21}(\pi/(2\Omega_{21}))U^{S}_{32}(\pi/(\sqrt{2}\Omega_{32}))$,
\begin{eqnarray}
&&-iU^{S}_{21}(\pi/(2\Omega_{21}))U^{S}_{32}(\pi/(\sqrt{2}\Omega_{32}))|T_2\rangle|\uparrow\rangle\nonumber\\
&&=-iU^{S}_{21}(\pi/(2\Omega_{21}))\frac{1}{\sqrt(2)}(|11\rangle|\uparrow\rangle -i|S_{3,0}\rangle|\downarrow\rangle)\nonumber\\
&&=-\frac{1}{\sqrt(2)}(|S_{21}\rangle|\downarrow\rangle +|S_{3,0}\rangle|\downarrow\rangle)=-|T_3\rangle|\downarrow\rangle.
\end{eqnarray}
The simulation of this process is shown in the Fig.~\ref{fig:figure7}.

\section{Dissipative dynamics}
In this section we study how the loss mechanisms affect the protocols introduced in the previous section. The master equation that describes the dissipative dynamics of our system (Eq.~(\ref{system})) is given by~\cite{PhysRevA.84.043832,PhysRevA.80.053810}
\begin{eqnarray}
\dot{\rho}&&=\frac{1}{i\hbar}[H,\rho]+\sum_{r=L,R}(U_{c_r}\rho S_{c_r}+S_{c_r}\rho U^{\dag}_{c_r}-S_{c_r}U_{c_r}\rho-\rho U^{\dag}_{c_r}S_{c_r})\nonumber\\
+&&\sum_{j=L,R,q}\sum_{m=x_j,z_j}(U_{m}\rho S_{m}+S_{m}\rho U^{\dag}_{m}-S_{m}U_{m}\rho-\rho U^{\dag}_{m}S_{m}),
\label{TCPOM}
\end{eqnarray}

where the operators $U_{\alpha}$ are defined as
\begin{eqnarray}
U_\alpha =&\int_0^{\infty}d\tau~\nu_\alpha(\tau)e^{-(i/\hbar) H\tau}S_\alpha e^{(i/\hbar) H\tau},\nonumber\\
\nu_\alpha(\tau)=&\int_{-\infty}^{\infty}d\omega~\frac{\gamma_\alpha(\omega)}{2\pi}[\bar{N}_\alpha(\omega)e^{i\omega\tau}+(\bar{N}_\alpha(\omega)+1)e^{-i\omega\tau}],\label{dissipation}
\end{eqnarray}

and we have considered transversal ($\gamma_x$) and longitudinal noise ($\gamma_z$) acting on two-level systems, and noise acting on the field quadrature ($\kappa$), through operators $S_{x_j}=\sigma^x_j$, $S_{z_j}=\sigma^z_j$, and $S_{c_r}=a^{\dag}_r+a_r$.

In particular, we show the effect of dissipation on the generation of $|S_{4,0}\rangle$ state. Before proceeding, it is instructive to estimate the times for carrying out the operations presented in section \ref{sec:IV}. Let us consider realistic values for circuit QED experiments that involve on-chip resonators coupled to flux \cite{Niemczyk2010} or transmon qubits \cite{Majer2007}. For instance, the resonator frequency on each QRS can be $\omega_c=2\pi\times 7$~GHz \cite{Houck2008}. In addition, we consider the QRS parameters $\omega_p=0.8\omega_c$, $g_p=0.5\omega_c$, that lead to matrix elements $|\chi_{10}|=1.1093$, $|\chi_{21}|=1.4916$, $|\chi_{42}|=1.7116$, $\Omega_\mu=0.004\omega_c$, and $g=0.002\omega_c$. In this case, we obtain the TLS rotation time $t_{TLS}$, $(1,0)$ SWAP interaction $t_{10}$, $(2,1)$ SWAP interaction $t_{21}$, and $(4,2)$ SWAP interaction $t_{42}$
\begin{eqnarray}
t_{TLS}=&&\left[\frac{\Omega_\mu}{2}|\langle\uparrow|\sigma^x|\downarrow\rangle|\right]^{-1}\frac{\pi}{2}\approx 17.86\,[{\rm ns}],\\ \nonumber
t_{10}=&&\left[\sqrt{2}g|\chi_{10}|\right]^{-1}\frac{\pi}{2}\approx 11.38\,[{\rm ns}],\\ \nonumber
t_{21}=&&\left[g|\chi_{21}|\right]^{-1}\frac{\pi}{2}\approx 11.97\,[{\rm ns}],\\ \nonumber
t_{42}=&&\left[g|\chi_{42}|\right]^{-1}\frac{\pi}{2}\approx 10.41\,[{\rm ns}],
\end{eqnarray}
that lead to a total time of about $87.34\,[{\rm ns}]$. 
\begin{figure}[t]
	\centering
	\includegraphics[scale=0.2]{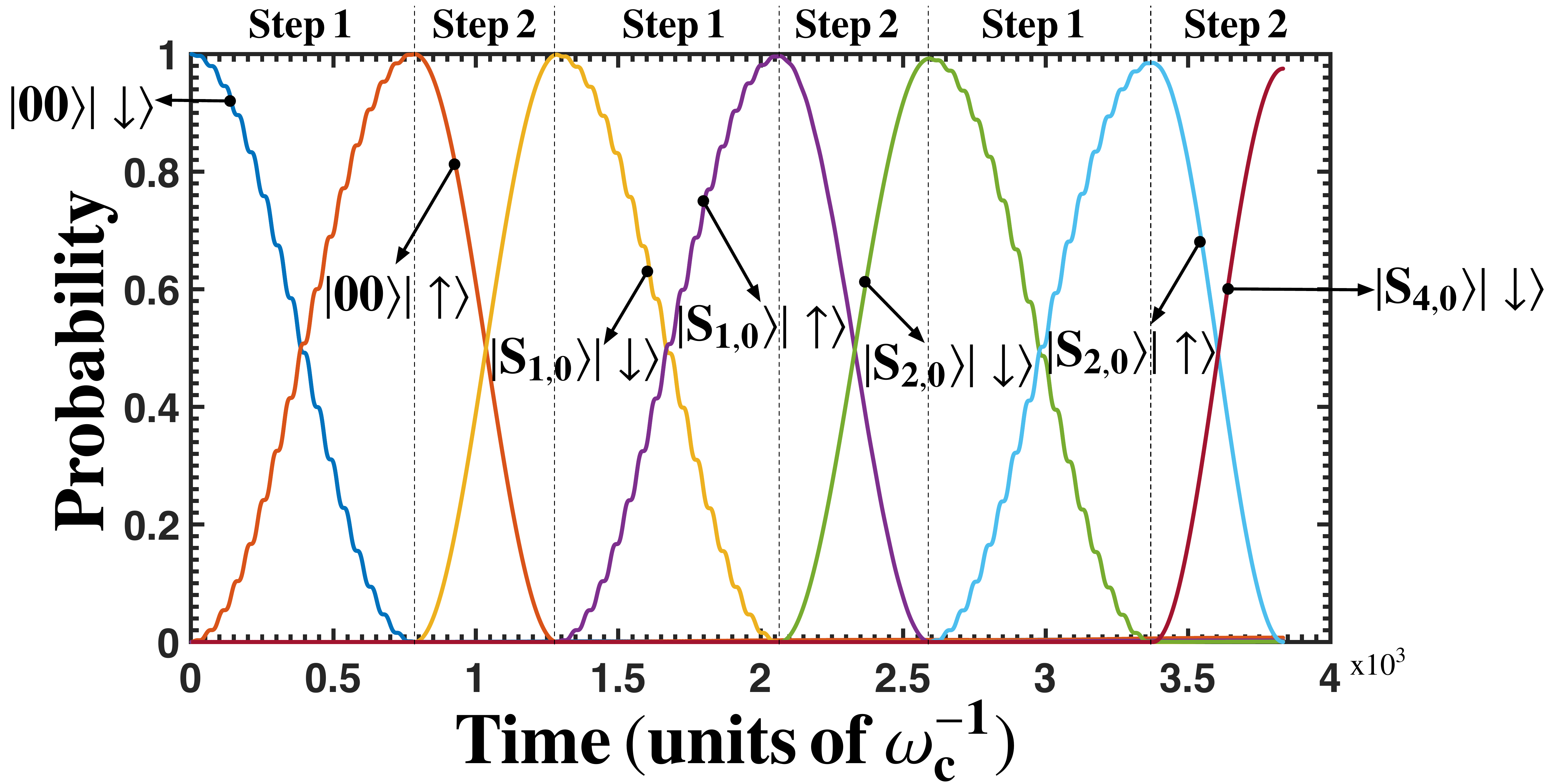}
	\caption{Generation of $|S_{4,0}\rangle$ state under dissipation with bare loss rates $\gamma^{-1}_x=4~[\mu s]$ and $\gamma^{-1}_z=0.2~[\mu s]$ for the qubits in the QRSs and ancilla TLS, and $\kappa=10~$[kHz] for each cavity.}
	\label{fig:figure8}
\end{figure}

We consider independent thermal baths at zero temperature, $\bar{N}_\alpha(\omega)=0$, for each loss mechanism and bare loss rates $\gamma^{-1}_x=4~[\mu s]$ and $\gamma^{-1}_z=0.2~[\mu s]$~\cite{Bertet2005} for qubits in the QRSs and ancilla TLS, and $\kappa=10$ kHz for each cavity \cite{Haack2010}, such that $\gamma_j(\omega)=(\gamma_j/\omega_j)\omega\Theta(\omega)$, where $\Theta(\omega)$ is the Heaviside step function \cite{Wang2016}. We have numerically tested that the protocol performance does not change if we consider a non-zero temperature bath within the operating temperature regime of circuit QED \cite{Clarke2008}. The step by step process to generate the state $|S_{4,0}\rangle$ including dissipation is shown in Fig.~\ref{fig:figure8}. The state is generated with fidelity $\mathcal{F}=0.9867$. Likewise, similar fidelities can be obtained for other protocols. For instance, in order to generate states $|S_{42}\rangle$, $|D_{42}\rangle$, and $\ket{T_3}$, we have obtained $\mathcal{F}=0.9771$, $\mathcal{F}=0.9705$, and $\mathcal{F}=0.9658$, respectively. 

We would like to stress that the dimensionality $N$ of the entangled states that we want to generate is mainly limited by loss mechanisms acting on the system. A high $N$ implies a large number of iterations in our protocols and a dynamics with larger decay rates according to Eq. (37), given by the different allowed energy transitions in our system. This in turn leads to effective decay times of about $1[\mu s]$ or less. Since the time for generating the state $|S_{N,0}\rangle$ with $N>4$ is larger than $0.1[\mu s]$, we have numerically tested that $N>4$ gives fidelities below $0.98$ for realistic circuit QED parameters, thus we obtain a bound $N=4$ for the dimension in our entangled states generation protocol.
\\
\section{Conclusions}
In summary, in this work we have studied a number of physical operations that allowed coherent manipulation in the Hilbert space of two QRSs, mediated by an ancilla TLS. In particular, we have applied such coherent operations for the generation of higher dimensional entangled states between QRSs. In addition we have considered the loss mechanisms acting on QRSs and ancilla TLS. The times involved for each operation are about $15~[{\rm ns}]$ when considering realistic parameters for circuit QED experiments, leading to fidelities larger than $95\%$. Finally, this proposal may have applications in circuit QED, within the context of ultrastrong coupling regime of light-matter interaction, such as parameter estimation for sensing magnetic fluxes in a quantum metrology approach, and the quantum simulation of complex spin systems such as Heisenberg interactions between high dimensional spins. 

\section*{Acknowledgements}
F.A.-A acknowledges support CONICYT Doctorado Nacional 21140432, G.A.B acknowledges support from CONICYT Doctorado Nacional 21140587, F.A.C.-L acknowledges support from  CEDENNA, basal grant No. FB0807 and Direcci{\'o}n de Postgrado USACH, G.R acknowledges the support from FONDECYT under grant No. 1150653, J. C. Retamal acknowledges the support from FONDECYT under grant No. 1140194.

\bibliographystyle{iopart-num}
\bibliography{references}
\end{document}